\documentclass[RNAAS]{aastex62}

\begin{document}

\title{On the central engine of the fastest-declining Type I supernova SN2019bkc}

\author{Shin'ichirou Yoshida}
\email{yoshida@ea.c.u-tokyol.ac.jp}
\affiliation{Department of Earth Science and Astronomy, Graduate School of Arts and Sciences,
The University of Tokyo\\
3-8-1 Komaba, Meguro-ku, Tokyo 153-18902, Japan}


\section{} 

\cite{Chen19} recently reported observations of a transient (SN 2019bkc/ATLAS19dqr)
which has been identified as a type I supernova with the shortest
declining time. Here I propose a simple model that explains the light curve of the transient
whose central engine is a remnant of non-explosive merger of double white dwarfs. The light curve
is modeled by the magnetic dipole radiation of the highly magnetized remnant.

Our model is based on the evolution scenario of a double dwarf merger described by \cite{Shen12}.
If a merger of a white dwarf binary resulting from gravitational wave radiation is not directly followed
by a violent merger explosion\citep{Pakmor12}, a highly differentially rotating remnant may remain
which evolves in viscous time scale $t=10^4 - 10^8$s after the merger, 
depending on the strength of the viscous mechanism. The viscosity induces exchanges of the angular momentum
inside the star
resulting in a slower and uniform rotation.
Meanwhile it heats and puffs the star up.
The merger may lead to the formation of a strong magnetic white dwarf \citep{Garcia-Berro12}. We assume
a strong magnetic field of order $10^9$G is established in the initial dynamical phase of the remnant.
In the viscous evolution phase, the virial theorem holds and the kinetic energy dissipation $\delta T$ 
into heat $\delta Q$ is
related to the increment of gravitational energy $\delta W$ and the internal energy $\delta U$ as
$
\delta W + \delta U = -2\delta T = 2\delta Q
$
, where we assumed adiabatic index $\gamma=4/3$. From the mass conservation, the radius
of the star is then $GM^2/R^2\dot{R} = \dot{Q}$ where the overdots represent time-derivatives.
We assume the viscosity is expressed by "alpha" prescription\citep{SS73} thus 
$
\dot{Q} = 4\pi \bar{c_s}/3\gamma \alpha^2\bar{p} R^2
$. Here $\bar{c_s}$ and $\bar{p}$
are averaged sound speed and pressure. Thus we have
$
\dot{R} = \beta R^4/(GM^2) \alpha^2\bar{c_s}\bar{p}
$
where $\alpha$ is the 'alpha' parameter of viscosity \citep{SS73} and $\beta$ is the factor of order unity.
By assuming hydrostatic equilibrium, we have $\bar{p}\sim GM\bar{\rho}\lambda/R$
and $\bar{c_s}\sim\sqrt{GM\lambda/R}$ where $\lambda$ is the ratio of average pressure
to the maximum and $\bar{\rho} = 3M/4\pi R^3$ . Finally we see how the radius changes as
\begin{equation}
\dot{R} = \kappa\alpha^2\sqrt{\frac{GM}{R}}
\label{eq: diff Rt}
\end{equation}
where $\kappa$ is a factor of the order of unity. The solution of Eq.(\ref{eq: diff Rt}) is
\begin{equation}
R(t) = R(0) \left(1+\alpha^2\kappa\sqrt{\frac{GM}{R(0)^3}} t\right)^{\frac{2}{3}}.
\label{eq: radius}
\end{equation}

Next we consider the central engine. 
The remnant conserves its total angular momentum in the viscous phase,
thus $R(t)^2\Omega(t) = R(0)^2\Omega(0)$. On the other hand from the
magnetic flux conservation we have $B(t)R(t)^2=B(0)R(0)^2$ where $B(t)$
is the surface magnetic field. Assuming the magnetic dipole radiation to be
the central engine, we have the luminosity$L(t)$  from the magnetic dipole
$\mu$ rotating at $\Omega$ as
\begin{eqnarray}
	L(t) &=& \frac{2\Omega(t)^4\mu(t)^2}{3c^3} = 4.4\times 10^{42}~(\rm{erg}~\rm{s}^{-1})~
	\left(\frac{\Omega(0)/2\pi}{1\rm{Hz}}\right)^4
	\left(\frac{R(0)}{0.01R_\odot}\right)^6
	\left(\frac{B(0)}{10^9\rm{G}}\right)^2
	\left(\frac{R(t)}{R(0)}\right)^{-6}\nonumber\\
	&=& 4.4\times 10^{42}~(\rm{erg}~\rm{s}^{-1})~
	\left(\frac{\Omega(0)/2\pi}{1\rm{Hz}}\right)^4
	\left(\frac{R(0)}{0.01R_\odot}\right)^6
	\left(\frac{B(0)}{10^9\rm{G}}\right)^2
	\left(1+\alpha^2\kappa\sqrt{\frac{GM}{R(0)^3}} t\right)^{-4},
	\label{eq: luminosity}
\end{eqnarray}
where we have used Eq.(\ref{eq: radius}).

\begin{figure}[h!]
\begin{center}
\includegraphics[scale=1.1,angle=0]{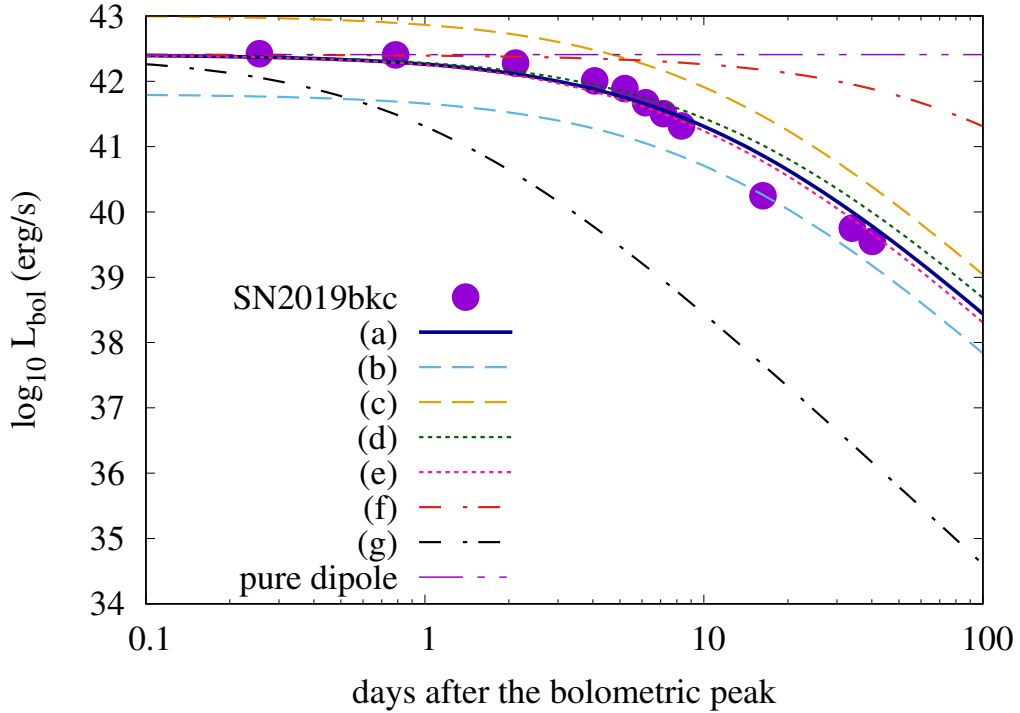}
\caption{Model bolometric light curves of Eq.(\ref{eq: luminosity}). (a): A best fit model of $M=1.1M_\odot$.
Model parameters are $(M/M_\odot,f_0,R(0)/R_\odot, B(0), \kappa\alpha^2)
=(1.1,0.53~\rm{Hz},0.014,9.4\times 10^8\rm {G},2.6\times 10^{-6})$. Here $f_0=\Omega(0)/2\pi$. 
(b): The same as (a) except $B(0)$ is halved.
(c): The same as (a) except $B(0)$ is doubled. (d): The same as (a) except $M=0.8M_\odot$.
(e): The same as (a) except $M=1.3M_\odot$. (f): The same as (a) except $\kappa\alpha^2$ is multiplied by $0.1$.
(g): The same as (a) except $\kappa\alpha^2$ is multiplied by $10$. The curve titled "pure dipole" is the case in which
pure magnetic dipole radiation is considered without the evolution of the remnant.
\label{fig:1}}
\end{center}
\end{figure}

In Fig.\ref{fig:1} we plot the light curve of Eq.(\ref{eq: luminosity}) as well as
the data points from Fig.4 of \cite{Chen19}. A model that reproduces 
the light curve well has a typical parameter of rapidly rotating 
(rotational frequency$\sim 0.5$Hz) and highly magnetized star ($B(0)\sim 10^9$G)
with the viscosity's $\alpha$ parameter to be ${\cal O}[10^{-3}]$ (since $\kappa$ is of order unity). 
Notice that the Kepler limit of spherical star is $f_0\sim 0.67$Hz for these parameter set.
With the higher (lower) magnetic field a star has a higher (lower) luminosity.
With the larger (smaller) $\alpha$ parameter the luminosity decreases more rapidly (slowly).
The dependence of the curves on the stellar mass is weak.
Finally, without the remnant's evolution being taken into account, the magnetic
dipole radiation has a too long decay time to explain the observed light curve.
The decay time scale $\tau$ of the radiation from a rotating magnetic dipole
with fixed strength is estimated as 
$
\tau~(\rm{d}) = \frac{3c^3I}{4m_\perp^2\Omega_0^2}\sim \frac{3c^3}{10}MR^{-4}B^{-2}\Omega_0^2\sim 3\times 10^7(\frac{M}{M_\odot})(\frac{R}{0.01R_\odot})^{-4}(\frac{B}{10^9G})^{-2}(\frac{f_0}{1Hz})^{-2}.
$

The central engine of our model is a rotating magnetic dipole, which may be reminiscent of 
a recently proposed engine for an X-ray transient powered by a newly-born magnetar \citep{Xue19}. 
There the spin-down luminosity is the energy source of a transient. In the current case, the decline 
of the luminosity cannot be explained solely by the spin-down of the fixed strength of dipole. In our model 
the rapid decline of luminosity results from the weakening of surface magnetic field
and from the spinning down of the star. Both of the factors are outcomes of the thermal expansion
of the star due to the viscous heating, whose energy source is the rapid differential rotation of the remnant.

\acknowledgments
This study is supported by JSPS Grant-in-Aid for Scientific Research(C) 18K03641.

\end{document}